\documentclass[conference]{IEEEtran}
\IEEEoverridecommandlockouts
\usepackage{cite}
\usepackage{amsmath,amssymb,amsfonts}
\usepackage{algorithmic}
\usepackage{graphicx}
\usepackage{textcomp}
\usepackage{xcolor}
\usepackage{subfigure}
\newcommand{\magenta}[1]{{\color{magenta}{#1}}}

\def\BibTeX{{\rm B\kern-.05em{\sc i\kern-.025em b}\kern-.08em
    T\kern-.1667em\lower.7ex\hbox{E}\kern-.125emX}}
\begin{document}

\title{Edge-Detect: Edge-centric Network Intrusion Detection using Deep Neural Network\\
\thanks{$\star$ Equal Contribution \newline\indent$\dagger$ Corresponding author: email: rmitra@semo.edu
}
}

\author{
Praneet Singh\textsuperscript{1$\star$},
Jishnu Jaykumar P\textsuperscript{1$\star$},
Akhil Pankaj\textsuperscript{1}, 
Reshmi Mitra\textsuperscript{2$\dagger$} \\
\textsuperscript{1} Indian Institute of Science, Bengaluru, India \\
\textsuperscript{2} Southeast Missouri State University, Cape Girardeau, USA \\
\small{\{praneetsingh, jishnuj, akhilpankaj\}@iisc.ac.in, rmitra@semo.edu}
}


\maketitle

\begin{abstract}
Edge nodes are crucial for detection against multitudes of cyber attacks on Internet-of-Things endpoints and is set to become part of a multi-billion industry. The resource constraints in this novel network infrastructure tier constricts the deployment of existing Network Intrusion Detection System with Deep Learning models (DLM).
We address this issue by developing a novel light, fast and accurate `Edge-Detect' model, which detects Distributed Denial of Service attack on edge nodes using DLM techniques. 
Our model can work within resource restrictions i.e. low power, memory and processing capabilities, to produce accurate results at a meaningful pace.
It is built by creating layers of Long Short-Term Memory or Gated Recurrent Unit based cells, which are known for their excellent representation of sequential data.
We designed a practical data science pipeline with Recurring Neural Network to learn from the network packet behavior in order to identify whether it is normal or attack-oriented. The model evaluation is from deployment on actual edge node represented by Raspberry Pi using current cybersecurity dataset (UNSW2015). Our results demonstrate that in comparison to conventional DLM techniques, our model maintains a high testing accuracy of $\sim$99\% even with lower resource utilization in terms of cpu and memory. In addition, it is nearly 3 times smaller in size than the state-of-art model and yet requires a much lower testing time.
\end{abstract}

\begin{IEEEkeywords}
Edge computing, Deep Learning, Internet of Things, DDoS, Recurrent Neural Networks
\end{IEEEkeywords}

\section{Introduction} \label{Sec:Introduction}
\noindent The Symantec 2019 Internet Security Threat Report~\cite{istr_2019} states that recently Internet of Things (IoT) has become a new infection vector for cyber attacks  such as Distributed Denial of Service (DDoS). 
However, attack detection is still an open and challenging problem because of dynamic, distributed, heterogeneous, and collaborative nature of the IoT devices. 
With their resource constraints, a growing body of work have postulated that threat detection function is best suited to be pushed to the edge nodes as first line of defense \cite{bhardwaj2018towards}. However, the resource-intensive algorithms of Deep Learning models (DLM) are unsuitable for the newly emerging network infrastructure tier i.e. edge nodes. 

Standard security datasets such as KDD Cup 99 \cite{stolfo2000cost}, MIT Lincoln Laboratory DARPA 2000 \cite{lippmann20001999}, CAIDA 2010 \cite{caida2010ddos}, and TUIDS DDoS dataset 2012 \cite{gogoi2012packet} are outdated with the advancement of computer network protocols and equipment.
According to best of our knowledge, a comprehensive dataset for the IoT or edge computing paradigm is currently nonexistent.  
From the state-of-art on attack detection using DLM, DeepDefense model \cite{yuan2017deepdefense} produced some of the best results in terms of prediction accuracy. Their efforts determined that Long Short-Term Memory (LSTM) and Gated Recurrent Unit (GRU) are the most effective DLM for analyzing network packets using the UNB ISCX Intrusion Detection Evaluation 2012 DataSet (now onwards ISCX2012 for short) \cite{shiravi2012toward}.

\begin{figure*}
   \centering
     \includegraphics[scale=.13]{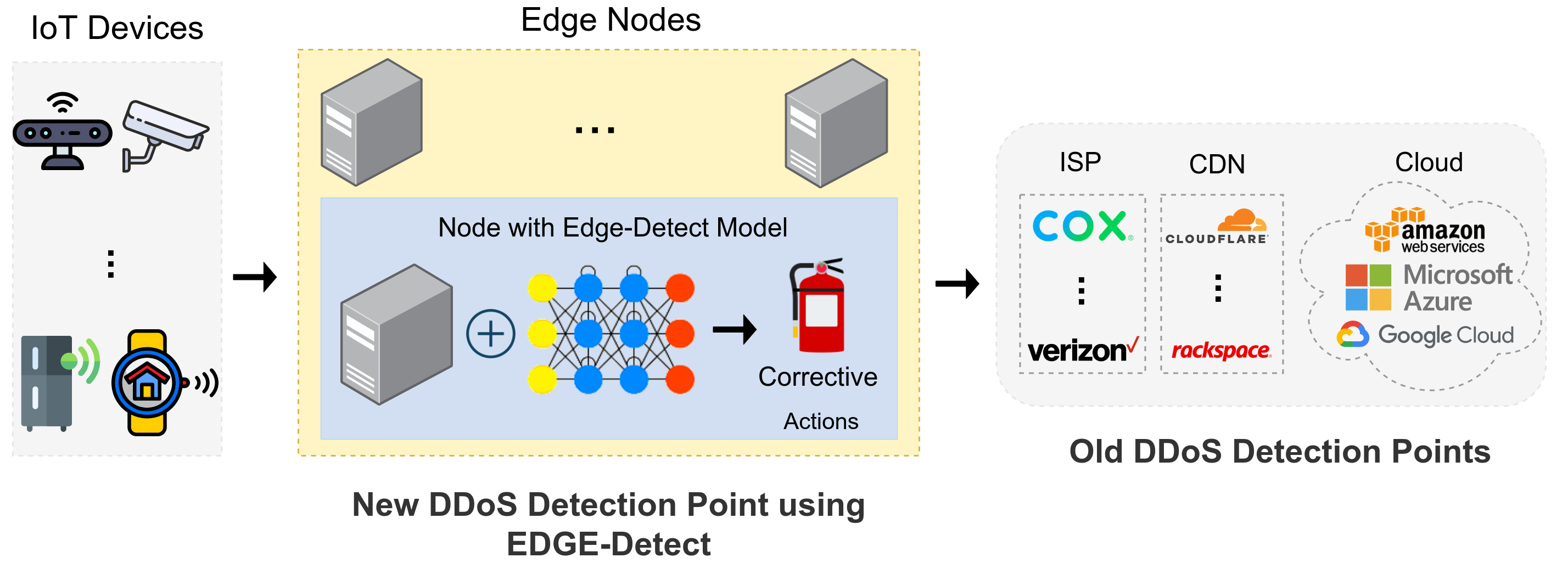}
\caption{Proposed location of Edge-Detect in comparison to the previous detection point in the cloud side. Transferring data from the point of attack up the hierarchy to cloud server causes crucial detection delays. Edge-Detect model (blue box) represents the edge node with added capabilities of our light and fast recurring neural network (RNN) module. The packet capture files from the IoT endpoints become the input, which are parsed and analyzed for detection purposes. This detection results can help drive the recommendation module for corrective actions.}
  \label{fig:edge-detect-location}
\end{figure*}

Motivated by these inadequacies, we have designed the \textbf{\textit{Edge-Detect}} model to enable DDoS detection on edge devices using light yet powerful and fast DLMs built by stacking the FAST cells~\cite{kusupati2018fastgrnn}. The term `light' emphasizes the resource requirements and `fast' denotes the processing performance. Edge-Detect is targeted towards the IoT security architect community, since the intended purpose is to safeguard IoT endpoints. Fig.~\ref{fig:edge-detect-location} brings to contrast the prior DDoS detection point located on the cloud server, which suffers from crucial detection latency. Being deployed at the edge node, our model becomes a faster path to examine the sequence of IoT network packets for potential attack. We chose UNSW2015 \cite{moustafa2015unsw} dataset for our evaluation, because it fulfilled all crucial criteria such as relatively current compared to other dataset, clear separation between training and testing sets, and correct labels for the network features.

The input for our DLM pipeline is a sequence of individual packets in the packet capture file, which are collated as windows with fixed length. This transformation is accompanied with reduced features and modification of attack label to signify whether an attack occurred in that particular window. 
These window sequences are processed through DLM which, in essence, is a network of LSTM or GRU cells. 
Our results verify that our model can outperform the state-of-art on multiple levels namely accuracy, precision, size (Kilobytes) and resource performance. To understand the model behaviour and deployment issues we are evaluating it on Raspberry Pi 3 and observed that other regular processes on the edge node are not starved.      
It maintains a high testing accuracy of $\sim$99\% even with lower resource utilization, and yet requires a much lower testing time.

Edge-Detect model is a significant improvement in the state-of-art because of few important reasons. First, the light FAST cells have not been used before to solve anomaly detection in network security problem in a low resource computing environment to the best of our knowledge.
Secondly, although several DLMs are available for IoT, limited details regarding model development, deployment and resource usage is available according to our literature review. 
Hence, it is very difficult to reuse and apply their work by IoT security architects with limited man hours or DLM skills. 
We are investigating practical design issues such as DLM network layers and feature engineering to build a realistic model and presented the validation results in this paper. 
In addition, our model is available in the public repository~\cite{rac_edd_github} and can be deployed by IoT security architect with minimal modification or training. 

The main contributions of this paper are listed below: 
\begin{enumerate}
\item Design a light and fast Deep Learning based Edge-Detect model by stacking layers of LSTM and GRU cells for resource-constrained computers. 
\item Model validation on a standard network security UNSW2015 dataset and conducted trade-off analysis on accuracy performance, and cost. 
\item Highlight accuracy and system performance issues by deploying our DLM on an actual edge node. 
\end{enumerate}

This paper has four main sections. In Section~\ref{Sec:related_work}, we are presenting the summary of state-of-art about DLM and cybersecurity analytics. Section~\ref{Sec:DLM} is about Edge-Detect model design. Model evaluation and discussion is part of Section~\ref{Sec:Evaluation}. We are concluding with the highlights of our work in Section~\ref{Sec:Conclusion}.

\section{Related Work} \label{Sec:related_work}

\noindent The earliest works on statistical techniques for intrusion detection in network packets \cite{anderson1995detecting} and~\cite{mauricio1998neural} appeared about two decades ago. The first comprehensive survey article~\cite{chandola2009anomaly} addressing the anomaly detection problem using machine learning techniques, appeared about a decade ago and called it ``Cyber-Intrusion Detection''. The techniques suggested in that article includes Bayesian Networks, Neural Networks, Support Vector Machines (SVM), Clustering and Nearest Neighbor. Many early challenges in applying  machine learning (ML) techniques for network intrusion detection system (NIDS) include understanding the threat model, keeping the model scope narrow and lack of training dataset among others as explained in ~\cite{sommer2010outside}. 

Early survey article~\cite{bhuyan2013network} to discuss specifically about the `network' intrusion detection within the anomaly detection area focused on the techniques, systems, datasets and tools related to NIDS. They have categorized them in terms of capability, performance, dataset used, matching and detection mechanisms, among others. A few key challenges presented in the paper include run-time limitation, dependence on the environment, nature of the anomaly and lack of unbiased dataset. Although DLM can solve some of these problems, but the need for a quality dataset which reflects the operating `symptoms' of an attack is still an unsolved issue.

RNN model comparison with different ML methods such as J48, artificial neural network, random forest, SVM for the NSL-KDD dataset \cite{nsl-kdd-dataset} is shown in \cite{yin2017deep}. The benchmark is an improved version of the earlier KDD dataset \cite{stolfo2000cost}. They have evaluated the impact on accuracy  w.r.t. classification (binary and multiclass), number of neurons and diverse learning rate reaching accuracy values of about 83\%.
We found out that traditional ML or shallow neural networks are impractical for large network traffic data which is set to reach the orders of zettabytes by 2021. In contrast, DLM eliminates the need for domain expertise by obtaining abstract correlations and reduces human efforts in pre-processing. 

DeepDefense model \cite{yuan2017deepdefense} is the earliest prominent work to use RNN based DLM for this problem domain. Their solution is based on RNN cells such as LSTM and GRU, because they have proved to be suitable with other sequential data problems such as speech recognition, language translation, speech synthesis among others. They have evaluated their DLM on the UNSB ISCX Intrusion Detection Evaluation 2012 dataset. 
We have advanced their work by building two different light and fast network models which are deployed on an actual edge node and achieving comparable accuracy margins on the low resource platform. 
Although several prior work exist for earlier datasets, we are ignoring them here for multiple reasons such as: datset relevance, less accuracy w.r.t. current models, inability to reproduce results which renders it unsuitable. 

Recent work \cite{roopak2019deep} has proposed DLM for the cybersecurity in IoT networks. The models provided very good detection accuracy of 97.16\%. In another excellent work \cite{zhang2019deep}, the authors have developed `deep hierarchical network' by cascading two types of networks (LeNet-5 and LSTM). They have applied their model on the CICIDS2017 \cite{sharafaldin2018toward} dataset, achieving an accuracy of about 90\%. However, they claim that their model can automatically select temporal and spatial features from existing input traffic without providing substantial details, which is a hard problem even in the ML community.

A prominent effort for social IoT \cite{diro2018distributed} is using distributed deep learning. Their model contains three hidden layers for feature learning and soft-max regression (SMR) for the classification task. 
In comparison to their distributed approach involving participating nodes exchanging multiple parameters which is computationally intensive, our focus is a centralized approach of maximizing the capabilities of each node. 
There have been other significant efforts in using DLM for attack detection in allied fields such as Software Defined Networks, such as \cite{niyaz2016deep}, \cite{li2018detection} and \cite{al2018deep}. However, we have limited our focus to IoT and edge devices for brevity.

\begin{figure}
    \centering
    \includegraphics[width=.48\textwidth]{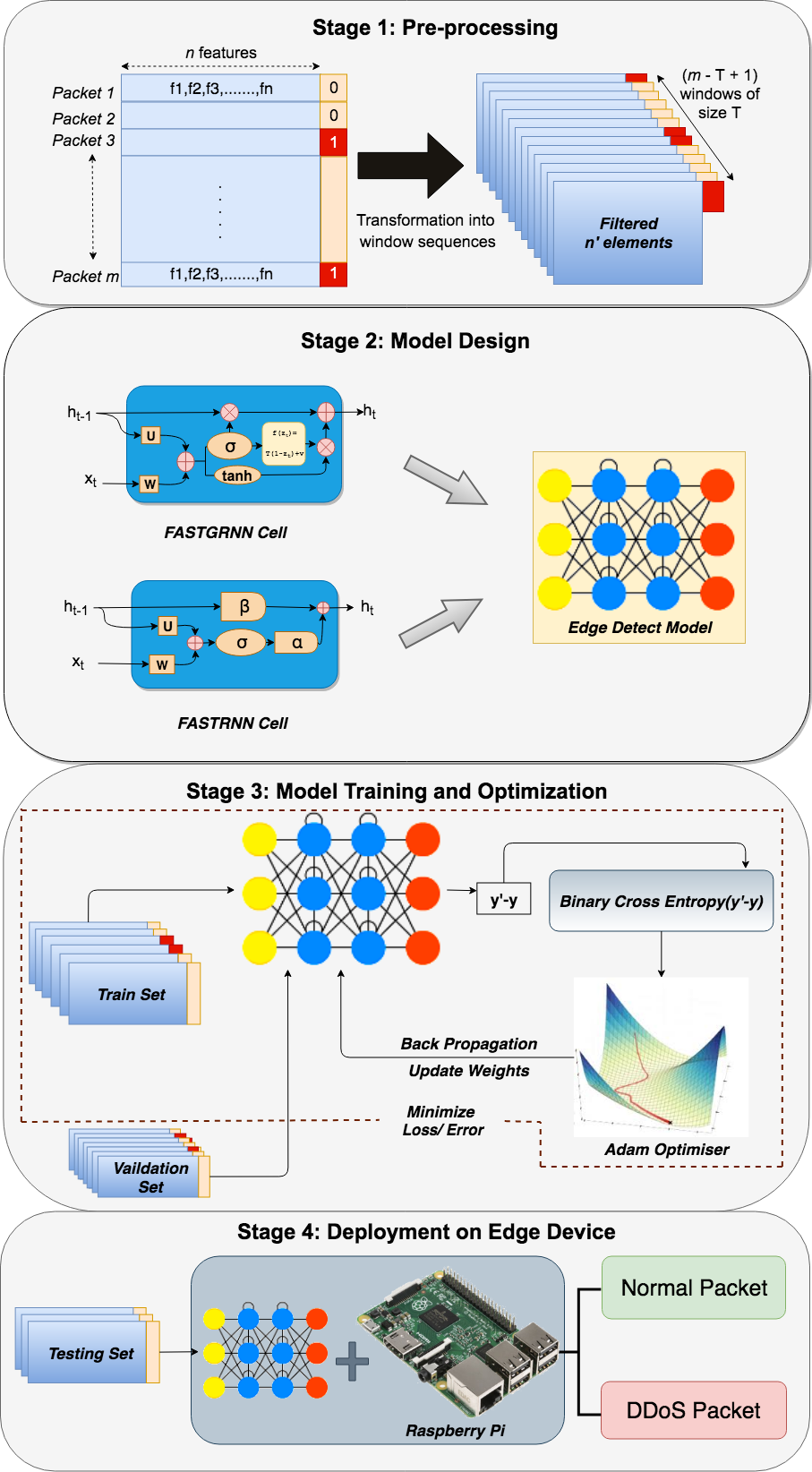}
    \caption{
    Edge DDoS detection pipeline consisting of these main steps: pre-processing, developing neural network model based on FAST cells, model training, optimization and testing on an edge device.}
    \label{fig:pipleine}
\end{figure}
\begin{figure*}
   \centering
  \includegraphics[scale=.35]{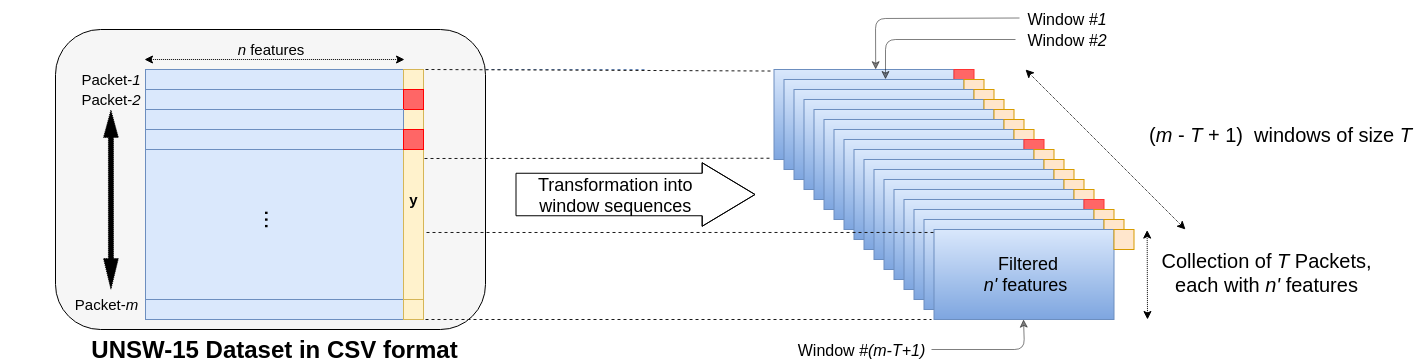}
  \caption{Packet to window transformation: In the Stage 2, input data from UNSW2015 dataset is transformed into window sequences using sliding window approach for RNN model training.}
  \label{fig:data_transformation}
\end{figure*}

\section{Edge DDoS Detection Pipeline} \label{Sec:DLM}

\noindent Built using standard data science techniques, the main pipeline stages in Edge-Detect are: (1) pre-processing, (2) neural network model design, (3) training and optimization, and (4) deployment on edge device. These stages are illustrated in Fig.~\ref{fig:pipleine}. 
The pre-processing stage uses the network packets from the UNSW2015 dataset to select significant features and convert individual packets into window sequences. 
Stage-2 involves DLM design using the provably fast and light FAST cells. They are suitable in comparison to the standard RNN cells such as LSTM and GRU, due to the residual gate connection \cite{kusupati2018fastgrnn}. 
Stage-3 in pipeline consists of training and optimization of RNN model  from Stage-2. 
Finally, the model is tested for deployement on an actual Raspberry Pi node to detect packets undergoing DDoS attack from the normal packets. 

\subsection{Stage 1: Pre-processing} \label{Sec:dataset_transformation}

\noindent UNSW2015 dataset is a network packet capture (pcap) CSV file, which is correctly pre-labeled as normal or attack-oriented. We are reducing it into a series of windows with reduced features, due to the limited processing capabilities of edge node. From the 49 features available in ICSX2012 (prior dataset), \cite{yuan2017deepdefense} produced remarkable results by applying DLM on the 20 features. 
To determine whether further reduction in the number of features is even possible, we used the 11 features of \cite{moustafa2017hybrid} as our reference. We are summarizing this feature selection process for the conciseness in this paper.
We performed a standard pre-processing technique called ``one hot encoding'' on the feature termed as ``state'' in the dataset to replace it by 15 additional features based on various category values. 

To build the comprehensive model about the network patterns in the entire dataset, it is important to  `learn' the characteristics from all preceding windows irrespective of the attack occurrence. 
This issue is addressed with the \textit{sliding window} approach, where each window is moved by a single packet to analyze whether the prior (\textit{T}-1) packets have led to an attack in the current packet.
This is shown in Fig.~\ref{fig:data_transformation}. 
A single window consisting of \textit{T} packets with $n$ features is reduced to $n'$  features with the binary label for the entire window depicting the occurrence of DDoS attack in the last packet. 
This means that applying DLM on this window is equivalent to learning from the information of its (\textit{T}-1) constituent packets to determine the attack occurrence in the \textit{T}-th packet. 
The $m$ packets produce a total of $(m - T + 1)$ windows due to these sliding windows.
Summarizing from the initial 49 ($= n$), the transformation in this stage yields to 25 ($= n'$) distinct features. 
Overall, at the end of this stage from the initial size of $(m \times n)$, the dataset at the end of this stage will have these $(m - T + 1)$ number of windows with each of them of the size $(T \times n')$. 

\subsection{Stage 2: Edge-Detect Model Design} \label{Sec:model_design}
\begin{figure}
    \centering
    \includegraphics[width=.35\textwidth, keepaspectratio]{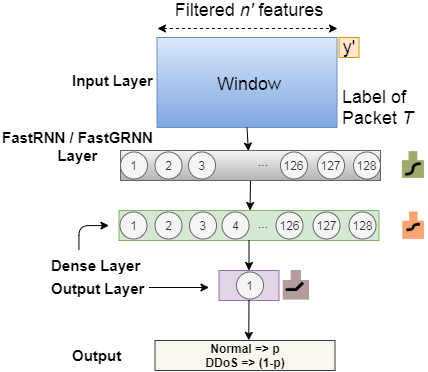}
        \label{fig:edd-rnn}
    \caption{Edge-Detect Model architecture. The output layer assigns probabilities $p$ and (1-$p$) to the input packet window for being normal and malicious respectively. Every RNN layer and every fully connected layer are followed by a batch normalization layer to accelerate network training}
    \label{fig:edd_network}
    \protect\includegraphics[height=0.35cm]{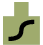} \footnotesize{ represents tanh,}
    \protect\includegraphics[height=0.35cm]{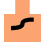}\footnotesize{ represents relu,}
    \protect\includegraphics[height=0.35cm]{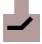}\footnotesize{ represents sigmoid.}
\end{figure}

\noindent   
%
Our model is built using layers of FAST cells which are either LSTM  or GRU. In fact, GRU is a variant of LSTM.
The advantage of using LSTM and GRU cells is that each unit `remembers' the existence of specific feature present in the input stream, which makes them successful for sequential applications. 
These LSTM/GRU layers are followed by a dense layer of 128 cells and finally the output layer as shown in Fig.~\ref{fig:edd_network}. 
The activation functions used are `tanh' for LSTM and GRU layers, `ReLU' for dense layer and `sigmoid' for output layer in all their models. 
We used the ReLU function as the activation function of the hidden layers. This is a non-linear activation function that can enhance the model performance by expressing a complicated classification boundary better than a linear activation function.

To signify whether it is associated with an attack, each packet is labeled in binary values of 1s and 0s in the (input) dataset.
This identification is inferred from probability values when applying the model. 
In our case, the output layer assigns certain probability values depending upon the weights learned from the previous layer including up to the dense layer.
In order to determine whether the packets are normal or attack-oriented, Edge-Detect model compares this probability with a certain threshold value. 
The results reported in the paper are with current value set as 0.8. 
The output layer is labeled as ``DDoS $=>$ (1-$p$)'' as shown in Fig.~\ref{fig:edd_network}.
\section{Model Evaluation} \label{Sec:Evaluation} 
\noindent We are evaluating the standard metrics such as accuracy, loss, precision and recall for the UNSW2015 dataset on edge node represented by Raspberry Pi 3, as represented by Stage-3 and 4 in Fig.~\ref{fig:pipleine}. The goal is to identify the most suitable DLM which can meet accuracy as well resource usage criteria for this newly emerging infrastructure tier. 
\subsection{Experimental Results}

\noindent The system-on-chip in our Raspberry Pi is Broadcom BCM2837 with quad-core ARM Cortex-A53 operating at frequency of 1.2GHz. In addition, it has a GPU of Broadcom VideoCore IV and RAM of 1GB LPDDR2 running at 900 MHz. For training, we are using Google Cloud services, where the CPU configuration is Intel Haswell with 8 virtual CPUS with 32 GB memory and GPU Tesla p100 with 100 GB HDD. We used Keras TensorFlow \cite{tensorflow2015-whitepaper} to enhance the FAST cells and, stack them in layers of LSTM and GRU for developing powerful networks with better accuracy. 

%
%

We begin our investigation by regenerating the results of DeepDefense model \cite{yuan2017deepdefense} using UNSW2015 dataset, since it was based on the prior ISCX2012 dataset. 
%
Although  our study began with the DeepDefense models, it is crucial to point out here that it is impossible to deploy them on the edge (Raspberry Pi) node. This is true even after feature engineering or scaling down the model by reducing number of cells/layers. 
During our preliminary investigation, we observed that these models are completely depleting the swap memory on this resource-constrainted platform. The authors have also concurred that building a light-weight model was not their intended purpose. 

\begin{table}[]
\caption{Performance evaluation of Edge-Detect results}
    \label{Table:results-edd}
    \centering
\begin{tabular}{|c|c|c|c|c|}
\hline
\textbf{Cell type}&\textbf{Accuracy}&\textbf{Loss}&\textbf{Precision}&\textbf{Recall} 
\\ \hline
FastRNN     & 99.6\%    & 4\%   & 99.5\%    & 99.75\%
\\ \hline
FastGRNN    & 99.5\%    & 2.4\% & 99.5\%    & 99.55\%
\\ \hline
\end{tabular}
\end{table}

\begin{table}[]
\caption{Comparing Edge-Detect with the  corresponding DeepDefense models}
    \label{Table:compare-dd-edd}
    \centering
\begin{tabular}{|c|c|c|c|c|c|}
\hline
\textbf{Category}&\textbf{Cell type}&\textbf{Weight}&\textbf{Accuracy}&\textbf{Layers}&\textbf{Cells}\\
\hline
DeepDefense & LSTM     & 1684 KB & 98\%  & 4     & 64  \\
\hline
DeepDefense & GRU      & 1314 KB & 98\%  & 4     & 64  \\
\hline
Edge-Detect  & FastRNN  & 598 KB & 99\%   & 1     & 128 \\
\hline
Edge-Detect  & FastGRNN & 609 KB & 99\%   & 1     & 128 \\
\hline
    \end{tabular}
\end{table}

\begin{table}[]
\caption{Results for Edge-Detect models built using cell types: FastRNN and FastGRNN. Both the model instances are single layer with 128 neurons. }
    \label{Table:compare-edd}
    \centering
\begin{tabular}{|c|c|c|}
\hline
\textbf{AUC}   & \textbf{KAPPA}  & \textbf{F1SCORE}  \\
\hline
99.96\% & 99.36\%  & 99.71 \\
\hline
\end{tabular}
\end{table}



Our model evaluation results for the key performance metrics namely accuracy, loss, precision and recall on the Raspberry Pi is summarized in Table~\ref{Table:results-edd}. The comparison with the state-of-art is shown in Table~\ref{Table:compare-dd-edd}.
%
In contrast to the high memory requirement (third column) of DeepDefense model, we have achieved a size reduction of 66\% with slightly better accuracy (fourth column). 
The weight drop is due to the adequacy of single layer in our model, whereas four layers requirement in their model. 
This is crucial for the platform resource restrictions, since it is impossible to accommodate large number of computations to achieve reasonable accuracy. 
This table also shows the cell types used for each model in the second column.
Table~\ref{Table:compare-edd} presents the AUC, Kappa and F1 score for our model results as represented by the last two rows of Table~\ref{Table:compare-dd-edd}. 
\newline

\begin{figure}
    \centering
    \subfigure[Hardware resource statistics from Raspberry Pi for the two variants of our Edge-Detect model built using FastRNN \& FastGRNN cells]
    {
        \includegraphics[width=.45\textwidth]{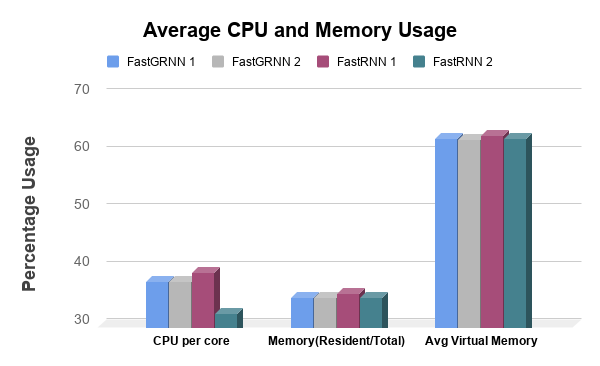}
        \label{fig:cpu_percentage_per_core}
    }
     
    \subfigure[Model testing time by executing Edge-Detect model on Raspberry Pi]
    {
        \includegraphics[width=.45\textwidth]{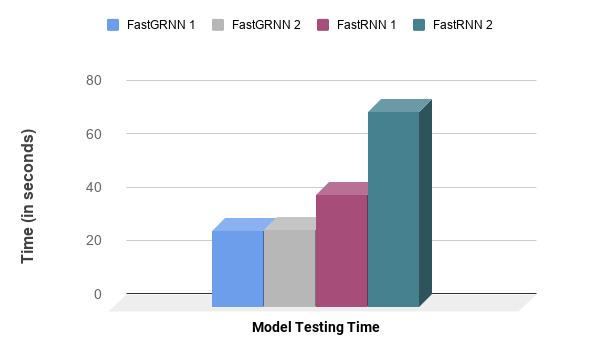}
        \label{fig:wall_clock_time}
    }

    \caption{Resource statistics to study the impact of Edge-Detect model on the memory and processing capabilities of the platform. Since both models have similar accuracy, resource usage is another dimension to explore for model selection.}
    \label{fig:pi_stats}
\end{figure}
\subsection{Resource Statistics}

\noindent Motivated with our accuracy results, we performed a deeper exploration to identify the most suitable and practical DLM for DDoS detection on the edge node.  
This second part of evaluation involves understanding the hardware execution parameters and trends. Using `top' (Linux utility), we are measuring the average utilization of cpu per core, resident and virtual memory in order to monitor the processor, swap memory and RAM while DLM is executing on the Raspberry Pi. We are using Linux `time' command to gauge the model testing time. These evaluations shown in Fig.~\ref{fig:pi_stats} demonstrate that comprehensively FastGRNN is slightly better suited for our cyber-defense application. Our experiments also revealed that Edge-Detect provides enough contigency for concurrent execution for other processes on the Raspberry Pi. We have done preliminary experimentation in this direction, however detailed investigation is beyond the scope of this paper.

\subsection{Discussion}
\noindent The \textit{first} main concept emerging out of the evaluation is that Edge-Detect has similar accuracy results ($\sim$99\%) for both the cell types FastRNN and FastGRNN. This observation is in alignment to the original FAST cell work \cite{kusupati2018fastgrnn}. 
\textit{Secondly}, windows with packets having fewer features is achieving comparable accuracy w.r.t. DeepDefense models on the newer dataset. 
By doing so,  we  reduce  the  number  of  required  computations on the edge node and  improve training time.  In addition, the experiments verify that testing accuracy increases after training the models with a selected feature dataset.
This is sustainable for the edge node, because loading windows with all the features is resource-intensive and hence, impractical for the Raspberry Pi. 
\textit{Thirdly}, high number of layers does not necessarily translate into better accuracy. For example, 64 neurons can produce similar accuracy and precision in a single layer as compared to four layers.  
\textit{Overall}, characteristics of Edge-Detect defies important processor and memory utilization concepts of prior DLM which has made it bulky for resource-constrainted platform. Our work also evolved towards the importance of feature engineering, which involves careful hyper-parameter tuning of learning rate, decay rate, batch size, among others. However, this is not the main objective of the paper and is beyond the current scope. 

\section{Conclusion} \label{Sec:Conclusion}
\noindent Deploying cyber-defense solutions based on standard NIDS techniques for IoT endpoints on network edge is a topic of immense current interest among academic and industry researchers. 
To analyze the exploding volumes of multi-modal network packet data, the most prevalent technique is anomaly detection based on ML/DLM. 
An appropriate deployment is closer to the IoT attack surface i.e. edge nodes. 
However, their minimal resources has imbalanced the trade-offs between prediction cost, deployment speed and accuracy. 
In order to overcome such limitations of the existing models, it is critical to develop new designs for resource-constrainted edge computing. 
In this paper, we are proposing Edge-Detect: a light and efficient DLM-enabled DDoS detection with edge nodes as deployment point.
Built using temporally sensitive neural networks such as LSTM and GRU, it learns from the network packet behavior to identify whether it is normal or attack-oriented. 
With minimal number of layers and light FAST cells, it can work within resource restriction to produce very accurate results with minimum training cost. 
We designed a practical data science pipeline based on RNN layers, validated it on a recent bulky UNSW2015 dataset and showed successful deployement. 
The investigation results demonstrate that in comparison to conventional DLM techniques, our model maintains a high testing accuracy of $\sim$99\%, while operating within the limited cpu resources such as memory, utilization and testing time.

Our future work involves developing a mitigation framework based on  Edge-Detect design. We also plan on automating the feature engineering and simplifying the training procedures of our model.
Secondly, we want to incorporate finer details of the packet features to make it more robust for a larger dataset. 
Finally, it is important to build a dataset which captures network behavior specifically for the IoT and edge networks.

\section{Acknowledgement}
\label{ACKNOWLEDGEMENT}
\begin{itemize}
    \item[$\ast$] Aditya Kusupati for his valuable suggestions.
    \item[$\ast$] Staff at IISc for providing us with their GPU and desktop to conduct our initial experiments.
\end{itemize}


\begin{thebibliography}{00}
\bibitem{istr_2019} ``Symantec Internet Security Threat Report'' vol. 24, February 2019 [Online]. Available https://docs.broadcom.com/doc/istr-24-2019-en [Accessed on July 21, 2020]
\bibitem{bhardwaj2018towards} K. Bhardwaj, J. C. Miranda, and A. Gavrilovska, ``Towards IoT-DDoS prevention using edge computing,'' in \textit{USENIX Workshop on Hot Topics in Edge Computing (HotEdge)}, 2018.
\bibitem{stolfo2000cost}  S.J. Stolfo, W. Fan, W. Lee, A. Prodromidis, and P.K. Chan, ``Cost-Based Modeling for Fraud and Intrusion Detection: Results from the JAM Project,'' in \textit{DARPA Information Survivability Conference and Exposition (DISCEX'00)}, pp. 130–144, 2000. 
\bibitem{lippmann20001999} R. Lippmann, J. W. Haines, D. J. Fried, J. Korba, and K. Das, ``The 1999 DARPA off-line intrusion detection evaluation,'' \textit{Elsevier Computer networks}, vol. 34, pp. 579–595, 2000.
\bibitem{caida2010ddos} Center for Applied Internet Data Analysis, UC San Diego, ``DDoS Attack 2007 dataset,'' 2010.
\bibitem{gogoi2012packet} P. Gogoi, M.H. Bhuyan, D.K. Bhattacharyya, and J.K. Kalita, ``Packet and flow based network intrusion dataset,'' in \textit{Springer International Conference on Contemporary Computing}, pp. 322–334, 2012.
\bibitem{yuan2017deepdefense} X. Yuan, C. Li, and X. Li, ``DeepDefense: Identifying DDoS Attack via Deep Learning,'' in \textit{IEEE International Conference on Smart Computing (SMARTCOMP)}, pp. 1-8, 2017.
\bibitem{shiravi2012toward} A. Shiravi, H. Shiravi, M. Tavallaee, and A.A. Ghorbani, ``Toward Developing a Systematic Approach to Generate Benchmark Datasets for Intrusion Detection,'' in \textit{Elsevier Computers \& Security}, vol. 31, pp. 357-374, 2012. 
\bibitem{kusupati2018fastgrnn} A. Kusupati, M. Singh, K. Bhatia, A. Kumar, P. Jain, and M. Varma, Manik, ``FASTgrnn: A Fast, Accurate, Stable and Tiny Kilobyte sized Gated Recurrent Neural Network,'' in \textit{Advances in Neural Information Processing Systems}, pp. 9017-9028, 2018.
\bibitem{moustafa2015unsw} N. Moustafa, and J. Slay, ``UNSW-NB15: A Comprehensive Data Set for Network Intrusion Detection Systems,'' in \textit{IEEE Military Communications and Information Systems Conference (MilCIS)}, pp. 1-6, 2015.
\bibitem{rac_edd_github} Edge-detect GitHub https://github.com/racsa-lab/EDD [Accessed on July 21, 2020].
\bibitem{anderson1995detecting} D. Anderson, T.F. Lunt, H. Javitz, A. Tamaru, and A. Valdes, ``Detecting Unusual Program Behavior Using the Statistical Component of the Next-generation Intrusion Detection Expert Systems (NIDES),'' Tech. Report SRI-CSL-95-06, SRI International. Computer Science Laboratory,  Menlo Park, California, 1995. 
\bibitem{mauricio1998neural} J. M. Bonif{\'a}cio,  A. M. Cansian, A. C. P. L. F. De Carvalho, and
E. S. Moreira, ``Neural Networks Applied in Intrusion Detection System,'' in \textit{IEEE World Congress on Computational Intelligence (WCCI)} pp. 205-210, 1998.
\bibitem{chandola2009anomaly} V. Chandola, A. Banerjee, and V. Kumar, ``Anomaly Detection: A Survey,'' in \textit{ACM Computing Surveys (CSUR)}, vol. 41, 2009.
\bibitem{sommer2010outside}  R. Sommer and V. Paxson, ``Outside the Closed World:
On Using Machine Learning For Network Intrusion Detection,'' in \textit{IEEE Symposium on Security and Privacy}, pp. 305–316, 2010.
\bibitem{bhuyan2013network} M.H. Bhuyan, D.K. Bhattacharyya, and J.K. Kalita, ``Network anomaly detection: Methods, systems and tools,'' \textit{IEEE Communications Surveys \& Tutorials}, vol. 16, pp. 303–336, 2013.
\bibitem{nsl-kdd-dataset} M. Tavallaee, E. Bagheri, W. Lu, and A. Ghorbani, ``A Detailed Analysis of the KDD CUP 99 Data set,'' in \textit{IEEE Symposium on Computational Intelligence for Security and Defence Applications}, 2009.
\bibitem{yin2017deep} C.L. Yin, Y.F. Zhu, J.L. Fei, and X.Z. He, ``A Deep Learning Approach for Intrusion Detection Using Recurrent Neural Networks,'' in \textit{IEEE Access}, vol. 5, pp. 21954–21961, 2017.
\bibitem{roopak2019deep} M. Roopak, G.Y. Tian and J. Chambers, ``Deep Learning Models for Cyber Security in IoT Networks,'' in \textit{IEEE Annual Computing and Communication Workshop and Conference (CCWC)}, pp. 452-457, 2019.
\bibitem{zhang2019deep} C. Zhang, P. Patras, and H. Haddadi, ``Deep Learning in Mobile and Wireless Networking: A Survey,'' in \textit{IEEE Communications Surveys \& Tutorials}, 2019. 
\bibitem{sharafaldin2018toward} I. Sharafaldin, A.H. Lashkari and A.A. Ghorbani, ``Toward Generating a New Intrusion Detection Dataset and Intrusion Traffic Characterization,'' in \textit{International Conference in Information System Security Privacy (ICISSP)}, pp. 108–116, 2018.
\bibitem{diro2018distributed} A.A. Diro and N. Chilamkurt,``Distributed Attack Detection Scheme using Deep Learning Approach for Internet of Things,'' in \textit{Elsevier Future Generation Computer Systems}, vol. 82, pp.761-768, 2017.
\bibitem{niyaz2016deep} Q. Niyaz, W. Sun, and A. Y. Javaid, A Deep Learning Based DDoS Detection System in Software-Defined Networking (SDN),'' , Submitted to EAI
Endorsed Transactions on Security and Safety, In Press, 2017, [Online].
Available: http://arxiv.org/abs/1611.07400 [Accessed on July 21, 2020]
\bibitem{li2018detection} C. Li, Y. Wu, X. Yuan, Z. Sun, W. Wang, X. Li, et al., ``Detection and Defense of DDoS Attack based on Deep Learning in Open Flow based SDN,'' in \textit{Journal of Communication Systems}, vol. 31, 2018.
\bibitem{al2018deep} M. Al-Qatf, Y. Lasheng, M. Al-Habib and K. Al-Sabahi, ``Deep Learning Approach Combining Sparse Autoencoder With SVM for Network Intrusion Detection,'' in \textit{IEEE Access}, vol. 6, pp. 52843-52856, 2018.
\bibitem{tensorflow2015-whitepaper} M. Abadi, A. Agarwal, et al, ``TensorFlow: Large-scale Machine Learning on Heterogeneous Systems,'' 2015, Software available from http:tensorflow.org [Accessed on July 21, 2020].
\bibitem{moustafa2017hybrid} N. Moustafa and J. Slay, ``A Hybrid Feature Selection for Network Intrusion Detection Systems: Central points, '' 2017, [online] Available: https://arxiv.org/abs/1707.05505 [Accessed on July 21, 2020].

\end{thebibliography}
\end{document}